\documentclass[aps,prb,twocolumn,superscriptaddress]{revtex4}
\usepackage{amsmath}
\usepackage{graphicx}
\usepackage[section]{placeins}

\newcommand{\ket}[1]{|#1\rangle}
\newcommand{\bra}[1]{\langle#1|}

\newcommand{\eq}{\begin{equation}}
\newcommand{\fine}{\end{equation}}
\newcommand{\eac}{\'e}

\begin{document}

\title{Optimal quantum cloning of orbital angular momentum photon qubits via Hong-Ou-Mandel coalescence}
\author{Eleonora Nagali}

\affiliation{Dipartimento di Fisica dell'Universit\`{a} ``La
Sapienza'', Roma 00185, Italy}

\author{Linda Sansoni}

\affiliation{Dipartimento di Fisica dell'Universit\`{a} ``La
Sapienza'', Roma 00185, Italy}

\author{Fabio Sciarrino}

\email{fabio.sciarrino@uniroma1.it}

\affiliation{Dipartimento di Fisica dell'Universit\`{a} ``La
Sapienza'', Roma 00185, Italy}

\affiliation{Istituto Nazionale di Ottica Applicata,
 Firenze 50125 , Italy}
\author{Francesco De Martini}

\affiliation{Dipartimento di Fisica dell'Universit\`{a} ``La
Sapienza'', Roma 00185, Italy}

\affiliation{Accademia Nazionale dei Lincei, via della Lungara 10,
Roma 00165, Italy}

\author{Lorenzo Marrucci}

\email{lorenzo.marrucci@na.infn.it}

\affiliation{Dipartimento di Scienze Fisiche, Universit\`{a} di
Napoli ``Federico II'', Compl.\ Univ.\ di Monte S. Angelo, 80126
Napoli, Italy}

\affiliation{CNR-INFM Coherentia, Compl.\ Univ.\ di Monte S. Angelo,
80126 Napoli, Italy}

\author{Bruno Piccirillo}

\affiliation{Dipartimento di Scienze Fisiche, Universit\`{a} di
Napoli ``Federico II'', Compl.\ Univ.\ di Monte S. Angelo, 80126
Napoli, Italy}

\affiliation{Consorzio Nazionale Interuniversitario per le Scienze
Fisiche della Materia, Napoli}

\author{Ebrahim Karimi}

\affiliation{Dipartimento di Scienze Fisiche, Universit\`{a} di
Napoli ``Federico II'', Compl.\ Univ.\ di Monte S. Angelo, 80126
Napoli, Italy}

\author{Enrico Santamato}

\affiliation{Dipartimento di Scienze Fisiche, Universit\`{a} di
Napoli ``Federico II'', Compl.\ Univ.\ di Monte S. Angelo, 80126
Napoli, Italy}

\affiliation{Consorzio Nazionale Interuniversitario per le Scienze
Fisiche della Materia, Napoli}

\begin{abstract}
\end{abstract}
\maketitle

\textbf{The orbital angular momentum (OAM) of light, associated with
a helical structure of the wavefunction, has a great potential for
quantum photonics, as it allows attaching a higher dimensional
quantum space to each photon\cite{Moli07,Fran08}. Hitherto, however,
the use of OAM has been hindered by its difficult manipulation. Here, exploiting the recently demonstrated spin-OAM information
transfer tools\cite{Marr06,Naga09}, we report the first observation
of the Hong-Ou-Mandel coalescence\cite{Hong87} of two incoming
photons having nonzero OAM into the same outgoing mode of a
beam-splitter. The coalescence can be switched on and off by varying
the input OAM state of the photons. Such effect has been then
exploited to carry out the $1\rightarrow2$ universal optimal quantum
cloning of OAM-encoded qubits\cite{Lama02,DeMa02,Scar05}, using the
symmetrization technique already developed for
polarization\cite{Ricc04,Irvi04}. These results are finally shown to
be scalable to quantum spaces of arbitrary dimension, even combining
different degrees of freedom of the photons.}


Since the OAM of photons lies in an infinitely dimensional Hilbert space, it provides a natural choice for implementing
single-photon \emph{qudits}, the units of quantum information in a
higher dimensional space. This can be an important practical advantage, as it allows increasing the information content per photon, and this, in turn, may cut down substantially the noise and losses arising from the imperfect generation and detection efficiency, by reducing the total number of photons needed in a given process. Qudit-based quantum
information protocols may also offer better theoretical performances
than their qubit equivalents\cite{Cerf02,Barb06}, while the combined
use of different degrees of freedom of a photon, such as OAM and
spin, enables the implementation of entirely new quantum
tasks\cite{Barr05,Barr08,Aoli07}. Finally, a OAM state of light can
be also regarded as an elementary form of optical image, so that OAM
manipulation is related to quantum image processing\cite{Gatt99}.

All these applications are presently hindered by the technical
difficulties associated with OAM manipulation. Despite important
successes, particularly in the generation and application of
OAM-entangled\cite{Mair01,Vazi03,Lang04} and OAM/polarization
hyper-entangled photons\cite{Barr05,Barr08}, a classic two-photon
quantum interference process such as the Hong-Ou-Mandel (HOM)
effect\cite{Hong87} has not been demonstrated yet with photons
carrying nonzero OAM. In the case of the polarization degree of
freedom, this phenomenon has played a crucial role in many recent
developments of quantum information, as well as in fundamental
studies of quantum nonlocality. It has been for example exploited
for the implementation of the quantum
teleportation\cite{Bouw97,Bosc98}, the construction of quantum logic
gates for quantum information processing\cite{Kok07}, the optimal
cloning of a quantum state\cite{Ricc04,Irvi04}, and various other
applications\cite{Kirc09}. Hitherto, none of these applications has
been demonstrated with OAM quantum states.

Quantum cloning -- making copies of unknown input quantum states --
represents a particularly important and interesting example. The
impossibility of making perfect copies, established by the
``no-cloning'' theorem\cite{Woot82}, is a fundamental piece of
modern quantum theory and guarantees the security of quantum
cryptography\cite{Gisi02}. Even though perfect cloning cannot be
realized, it is still possible to single out a complete positive map
which yields an \textit{optimal quantum cloning}\cite{Scar05}
working for any input state, that is, \textit{universal}. By this
map, an arbitrary, unknown quantum state can be experimentally
copied, but only with a cloning fidelity $F$ -- the overlap between
the copy and the original quantum state -- lower than unity.
Implementing quantum cloning is useful whenever there is the need to
distribute quantum information among several parties. The concept
finds application also in the security assessment of quantum
cryptography, the realization of minimal disturbance measurements,
in enhancing the transmission fidelity over a lossy quantum channel,
and in separating the classical and quantum
information\cite{Scar05,Ricc05}. Optimal quantum cloning machines,
although working probabilistically, have been demonstrated
experimentally for polarization-encoded photon qubits by stimulated
emission\cite{Lama02,DeMa02} and by the symmetrization
technique\cite{Ricc04,Irvi04,Sciar04}. In the latter method, the
bosonic nature of photons, i.e.\ the symmetry of their overall
wavefunction, is exploited within a two-photon HOM coalescence
effect. In this process, two photons impinging simultaneously on a
beam splitter (BS) from two different input modes have an enhanced
probability of emerging along the same output mode, i.e.\ of
coalescing, as long as they are undistinguishable. If the two
photons are made distinguishable by their internal quantum state,
for example encoded in the polarization $\pi$ or in other degrees of
freedom, the coalescence effect vanishes. Now, if one of the two
photons involved in the process is in a given input state to be
cloned and the other in a random one, the HOM effect will enhance
the probability that the two photons emerge from the BS with the
same quantum state, i.e.\ with successful cloning, when they emerge
together along the same output mode of the BS. For qubit states, the
ideal success probability of this scheme is $p=3/4$ (when exploiting
both BS exit ports), while the cloning fidelity for successful
events is $F=5/6$, corresponding to the optimal value\cite{Sciar04}.
The probabilistic feature of this implementation does not spoil its
optimality, since it has been proved that the optimal cloning
fidelity is the same for any probabilistic procedure\cite{Fiur04}.

In this paper, we report the first observation of two-photon HOM
coalescence interference of photons carrying nonzero OAM.
Furthermore, we exploit this result to demonstrate for the first
time the $1\rightarrow2$ universal optimal quantum cloning (UOQC) of
the OAM quantum state of a single photon. More specifically, we show
that we can optimally clone any qubit state $|\varphi\rangle_{o_2}$
encoded in the photon OAM bidimensional subspace $o_2$, spanned by
the eigenstates $\{|+2\rangle,|-2\rangle\}$ respectively
corresponding to an OAM of $+2$ and $-2$, in units of $\hbar$. A key
technical progress which made these results possible is given by the
polarization-OAM bidirectional quantum transfer devices that we have
recently demonstrated\cite{Naga09}, whose working principle is based
on the spin-to-orbital optical angular momentum conversion process
taking place in the so-called \textit{q-plates}\cite{Marr06}.

\begin{figure}[h]
\centering
\includegraphics[width=8cm]{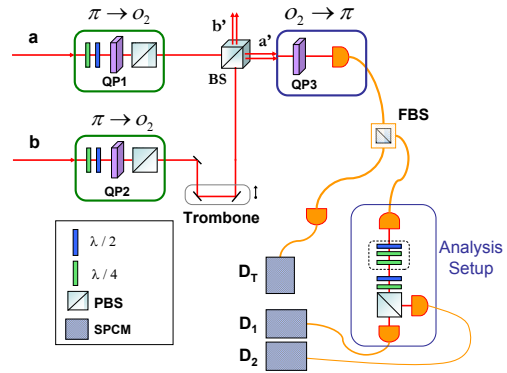}
\caption{\textbf{Experimental setup for demonstrating the Hong-Ou-Mandel
effect and for implementing the OAM quantum
cloning.} In the experiments, two photons generated by parametric
fluorescence and having the same wavefunction (see Methods for more
details) are injected in the two input modes $a$ and $b$. To set the
input OAM state $|\varphi\rangle_{o_2}$ of a photon, the same state
is first encoded in the polarization space $\pi$, i.e.\ the state
$|\varphi\rangle_{\pi}$ is prepared by using a combination of a
quarter-wave plate ($\lambda/4$) and a half-wave plate
($\lambda/2$), and then the $\pi\rightarrow o_2$ quantum transfer
device\cite{Naga09} is exploited. In particular, a transferrer
composed of a q-plate (QP1 in mode $a$ and QP2 in $b$) and a
polarizer (PBS) filtering the horizontal ($H$) polarization achieves
the transformation $|\varphi\rangle_{\pi}|0\rangle_{o} \Rightarrow
|H\rangle_{\pi}|\varphi\rangle_{o_2} $, where $|0\rangle_o$ denotes
the zero-OAM state. By this method, the input OAM state of each of
the input photons was set independently. The photons in modes $a$
and $b$ then, after a time synchronization controlled by the
trombone device, interfere in the balanced beam splitter (BS),
giving rise to the HOM effect and/or to the cloning process in the
BS output mode $a^{\prime}$. In order to analyze the OAM quantum
state of the outgoing photons, a $o_2\rightarrow\pi$
transferrer\cite{Naga09} is first used. This device, combining a
q-plate and a coupler into a single-mode fiber, achieves the inverse
transformation $|H\rangle_{\pi}|\varphi\rangle_{o_2} \Rightarrow
|\varphi\rangle_{\pi}|0\rangle_{o}$, thus transferring the quantum
information contained in the two photons back into the polarization
degree of freedom, where it can be easily read-out. The two photons
coupled in the single mode fiber are then separated by a fiber
integrated BS (FBS) and detected in coincidence, after analyzing the
polarization of one of them by a standard analysis setup. The signal
is given by the coincidences between single-photon counting module
(SPCM) detectors $D_T$ and either $D_1$ or $D_2$, thus corresponding
to a post-selection of the sole cases in which both photons emerge
from the BS in mode $a^{\prime}$ and are then split by the FBS. } \label{setup}
\end{figure}

As we work in a bidimensional subspace of the orbital angular
momentum, it is possible to construct a ``Poincar{\'e}'' (or Bloch)
sphere for representing the state of an OAM qubit that is fully
analogous to the one usually constructed for a polarization
qubit\cite{Padg99}. Being $\{|+2\rangle,|-2\rangle\}$ the basis in
the OAM subspace $o_2$, which can be considered the OAM equivalent
of the circular polarization states, we may introduce the following
superposition states
$|h\rangle=\frac{1}{\sqrt{2}}(|+2\rangle+|-2\rangle)$, $|v\rangle=
\frac{1}{i\sqrt{2}}(|+2\rangle-|-2\rangle)$,
$|a\rangle=\frac{1}{\sqrt{2}} (|h\rangle+|v\rangle)$, and
$|d\rangle=\frac{1}{\sqrt{2}}(|h\rangle-|v\rangle)$, which are the
OAM equivalent of horizontal/vertical and antidiagonal/diagonal
linear polarizations. The OAM eigenstates $\ket{+2},\ket{-2}$ have
the azimuthal transverse pattern of the Laguerre-Gauss (LG) modes,
while states $\ket{h},\ket{v}$ (as well as $\ket{a},\ket{d}$) have
the azimuthal structure of the Hermite-Gauss (HG) modes.

As a first experimental step, we carried out a HOM coalescence
enhancement measurement. To this purpose, we prepared the two input
photons in a given input OAM degree of freedom (see Fig.\
\ref{setup} for details). The time delay controller device and the input
polarizers guarantee also the perfect temporal and polarization
matching of the two photons, so as to make them undistinguishable,
except possibly for the OAM state. We expect to observe the
constructive interference between the two photons only if the OAM
contribution to the bosonic wavefunction is also symmetric in the
output, that is if the two outgoing photons have the same OAM. Let
us note that the BS is not an OAM preserving optical device, as the
reflection on the BS inverts the OAM sign. Therefore, the maximal
two-photon coalescence is expected to be observed when the input
photons carry exactly opposite OAM. Moreover, coalescence is
expected when they have the same HG-like state, e.g. $\ket{h}$ or
$\ket{v}$.
\begin{figure}[h]
\centering
\includegraphics[scale=.26]{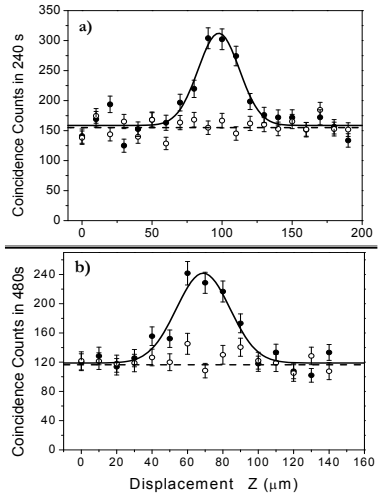}
\caption{\textbf{Experimental Hong-Ou-Mandel effect with OAM.} \textbf{a})
Input photons having opposite-OAM eigenstates $\ket{+2}$ and
$\ket{-2}$ (full circles) and equal-OAM eigenstates $\ket{+2}$ (open
circles). The first case leads to coalescence enhancement when the
photons are synchronized by the trombone displacement, while the
second case shows no enhancement. The solid curve is a best fit
based on theoretical prediction. \textbf{b}) Input photons having
the same HG-like OAM-superposition state $\ket{h}$ (full circles)
and orthogonal HG-like states $\ket{h}$ and $\ket{v}$ (open
circles). The error bars assume a Poisson distribution and are set
at one standard deviation.} \label{hom}
\end{figure}
Fig.\ \ref{hom} shows the results of our experiments. As expected,
we find a peak in the coincidence counts when the photon arrivals on
the beam-splitter are synchronized and the OAM output states are
identical. When the OAM output states are orthogonal the coalescence
effect is fully cancelled. The mean coincidence-counts enhancement
observed in the two Hong-Ou-Mandel experiments is $R=(1.97\pm
0.05)$, in agreement with the theoretical value of 2.

We now move on to the OAM cloning experiment. Let us first briefly
describe the theory of the UOQC process in the OAM subspace $o_2$.
The OAM qubit to be cloned, $|\varphi\rangle_{o_{2}}=\alpha
|+2\rangle +\beta |-2\rangle$, is attributed to the photon in input
mode $a$. The photon in input mode $b$ is prepared in the mixed
state described by density matrix $\rho_{o_{2}}^{b}=
\left(|+2\rangle \langle +2|+|-2\rangle \langle -2|\right)/2$. The
two photons are then made to interfere in the BS. By selecting the
cases in which the two photons emerge from the BS in the same output
mode $a^{\prime }$, the overall two-photon state is then subject to
the following projection operator: $P^{a^{\prime}}=\left( |\Psi
^{+}\rangle^{a^{\prime }}\langle \Phi ^{+}|^{ab}+|\Phi ^{+}\rangle
^{a^{\prime }}\langle \Psi ^{+}|^{ab}+|\Phi ^{-}\rangle^{a^{\prime
}}\langle \Psi ^{-}|^{ab}\right) $\ with $|\Phi ^{\pm }\rangle
=2^{-\frac{1}{2}}(|+2\rangle |+2\rangle \pm |-2\rangle |-2\rangle )$
and $|\Psi ^{\pm}\rangle=2^{-\frac{1}{2}}(|+2\rangle |-2\rangle \pm
|+2\rangle |-2\rangle )$. Note that this projection operator takes
into account the fact that the photon OAM undergoes a sign inversion
on reflection in the BS. The two photons emerging in mode
$a^{\prime}$ are then separated by means of a second beam splitter
and each of them will be cast in the same mixed qubit state
\begin{equation}
\rho _{o_{2}}^{a^{\prime }}=\frac{5}{6} |\varphi \rangle
_{o_{2}}\langle \varphi |+\frac{1}{6}|\varphi ^{\bot }\rangle
_{o_{2}}\langle \varphi ^{\bot }|
\end{equation}
which represents the optimal output of the $1\rightarrow 2$ cloning
process of the state $|\varphi \rangle _{o_{2}}$, with fidelity
$F=\;_{o_{2}}\langle \varphi |\rho _{o_{2}}^{a\prime}|\varphi
\rangle _{o_{2}}=\frac{5}{6}$.

\begin{figure}[t]
\centering
\includegraphics[scale=.22]{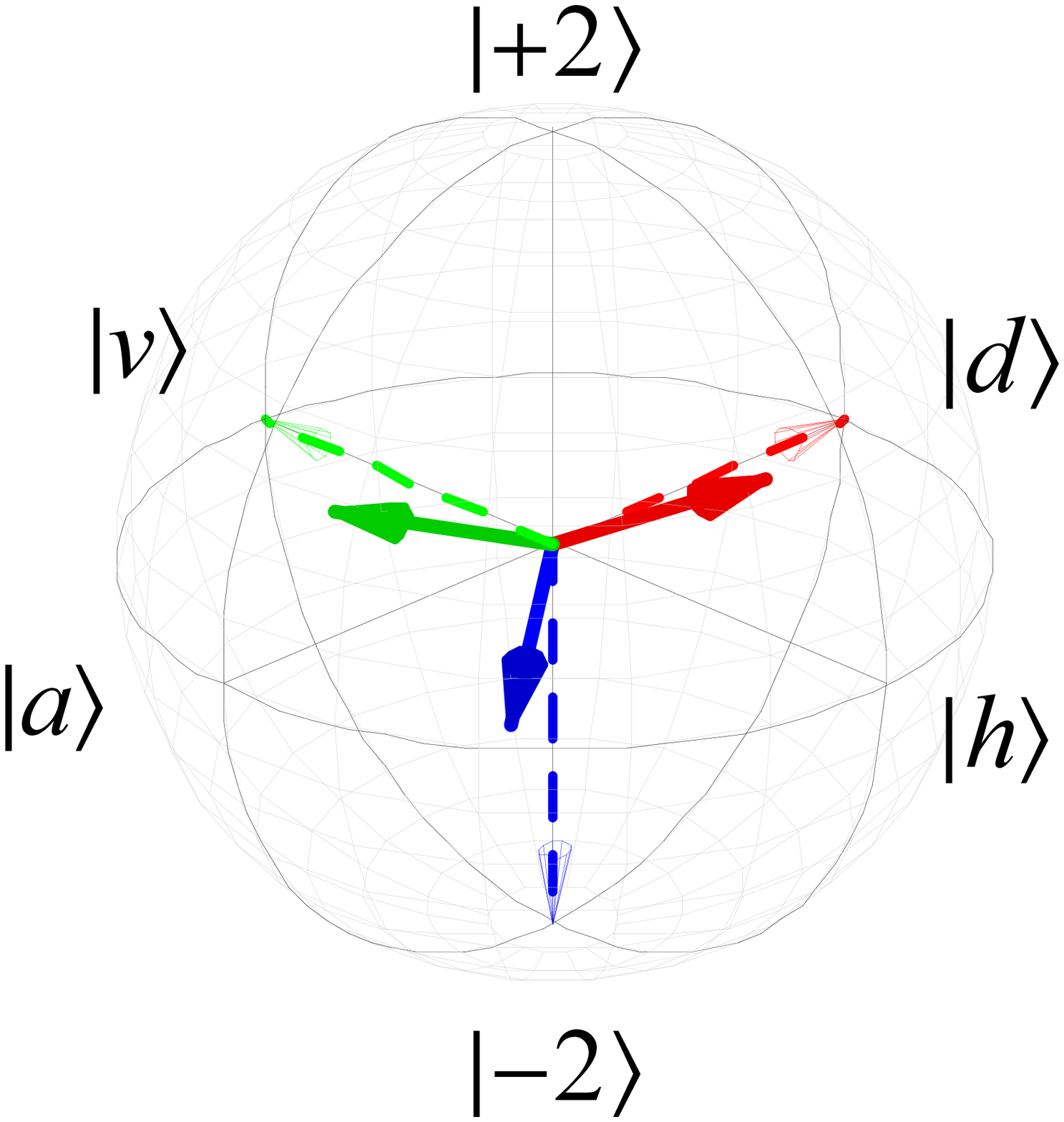}
\caption{\textbf{Experimental shrunk Bloch sphere of the OAM cloned qubits
in the subspace $o_2$.} The dashed-line arrows refer to the
theoretical input qubits, while the solid-line arrows give the
experimental output ones.} \label{sphere}
\end{figure}

Experimentally, the mixed state $\rho_{o_{2}}^{b}$ has been prepared
by randomly rotating, during each experimental run, a half-wave
plate inserted before the q-plate QP2 (see Fig.\ref{setup}).
\begin{table}[h]
\begin{center}
\begin{tabular}{||c||c||}
\hline
$\text{State}$ & $\text{Fidelity}$ \\ \hline
$|h\rangle_{o_2}$ & $(0.806\pm0.023)$ \\ \hline
$|v\rangle_{o_2}$ & $(0.835\pm0.015)$ \\ \hline
$|-2\rangle_{o}$ & $(0.792\pm0.024)$ \\ \hline
$|+2\rangle_{o}$ & $(0.769\pm0.022)$ \\ \hline
$|a\rangle_{o_2}$ & $(0.773\pm0.020)$ \\ \hline
$|d\rangle_{o_2}$ & $(0.844\pm0.019)$ \\ \hline\hline
\end{tabular}
\\[0pt]
\end{center}
\par
\caption{\textbf{Experimental fidelities for the cloning process.} The experimental values of the fidelity are reported for six specific OAM states.}
\end{table}
In Table I, we report the experimental fidelities of the cloning
process for six different input states $|\varphi \rangle _{o_{2}}$
to verify the universality of the cloning process(see Methods for further details). The measured values are in good
agreement with the theoretical prediction $F=\frac{5}{6}$ (see Methods). For the sake of completeness,
we have also measured the four Stokes parameters\cite{Padg99} of
some cloned states. The results are reported in Fig.\ \ref{sphere}.
Experimentally, the mean length of the vectors on the Bloch's sphere
representing the cloned states is found to be $S_{exp}=(0.68\pm
0.02)$, to be compared with the theoretical value $S_{th}=2F-1=2/3$.
The value of $S_{exp}$ has been estimated as
$S_{exp}=\sqrt{S_{1}^{2}+S_{2}^{2}+S_{3}^{2}}$, where $S_{i}$ refers
to the $i$-th measured Stokes component on the Bloch's sphere.
Hence, the optimal cloning process corresponds to a shrinking of the
whole Bloch sphere, in the subspace $o_{2}$: the unitary vector
length, related to the visibility of the input qubit, is shortened
to two thirds in the output.

Finally, as shown in the Methods section, we note that the
symmetrization technique that implements the quantum cloning is
optimal not only for qubit states, but also for arbitrary dimension
$d$ of the internal spaces of the quantum systems that are cloned
(qudits): photons with internal states defined in arbitrarily large
subspaces of OAM, or even spanning different internal degrees of
freedom at the same time (e.g., polarization, time, arbitrary
transverse modes, etc.), can also be cloned optimally by the same
method. The fidelity will be given by
$F=\frac{1}{2}+\frac{1}{d+1}$,\cite{Nave03} while the success
probability decreases only weakly with increasing $d$, saturating at
$p=1/2$ in the $d\rightarrow\infty$ limit.

In conclusion, in this paper we have experimentally carried out the
first observation of a two-photon Hong-Ou-Mandel coalescence of
photons carrying nonzero OAM and the universal optimal quantum
cloning of OAM qubits. These results open the way to the use of OAM
in many other quantum information protocols based on two-photon
interference effects, such as the generation of complex entangled
states (e.g., cluster states), purification processes, and the
implementation of logic gates.

\section*{Methods}
\begin{footnotesize}
\textbf{Experimental setup.}
The input photon pairs are generated via spontaneous parametric
fluorescence in a $\beta$-barium borate crystal, pumped by the second harmonic of a Ti:Sa
mode-locked laser beam. The generated photons have horizontal ($H$)
and vertical ($V$) linear polarizations, wavelength $\lambda=795$
nm, and spectral bandwidth $\Delta\lambda=6$ nm, as determined by an
interference filter. The detected coincidence rate of the source is
$C_{source}=5$ kHz. The photons are delivered
to the setup via a single-mode fiber, thus defining their transverse
spatial mode to a TEM$_{00}$. The photons are then split by a
polarizing beam splitter and injected in the two input modes $a$ and
$b$ of the apparatus shown in Fig.\ \ref{setup}. The
quantum transferrers $\pi\rightarrow o_2$ inserted on the modes $a$
and $b$ are based on q-plates having topological charge $q=1$,
giving rise to the OAM conversion $m \rightarrow m\pm2$ (in units of
$\hbar$) in the light beam crossing it, where the $\pm$ sign is
fixed by the input light circular-polarization
handedness\cite{Marr06,Naga09}. The conversion efficiency of the
q-plates was $0.80\pm0.05$ at $795$ nm, due to the reflection on the
two faces, not-perfect tuning of the q-plate and birefringence
pattern imperfections. The overall transferrer fidelity within the
output OAM subspace is estimated at $F_{prep}=(0.96\pm 0.01)$,
mainly due to the imperfect mode quality of the q-plates, leading to
a non-perfect $\pi\rightarrow o_2$ conversion. The polarization and
temporal matching on the beam splitter between photons on mode $a$
and $b$ has been achieved within an estimated error fixed by the
polarization setting accuracy ($0.5^{\circ}$) and the positioning
sensitivity of the trombone device ($1\mu$m).\\

\textbf{Cloning fidelity and success rate estimation.}
We set the polarization analysis system so as to have detectors
$D_{1}$ and $D_{2}$ (see Fig.\ \ref{setup}) measuring photons
respectively in the cloned state $|\varphi \rangle _{o_{2}}$ and in
the orthogonal one $|\varphi^{\bot }\rangle _{o_{2}}$. The
coincidences of either one of these detectors with $D_{T}$ ensure
also the coalescence of the two photons in the same mode. Let
$C_{1},C_{2}$ denote the coincidence counts of $D_{T}$ and
$D_{1},D_{2}$, respectively. The cloning success rate is then
proportional to $C_{tot}=C_1+C_2$, while the average experimental
cloning fidelity is given by $F_{exp}=C_{1}/C_{tot}$.

The experimental cloning fidelity is to be compared with the
prediction that takes into account the imperfect preparation
fidelity $F_{prep}$ of the OAM photon state to be cloned (the
fidelity of the mixed state is higher than 0.99, due to
compensations in the randomization procedure), given by
$F_{th}=\frac{F_{prep}R+\frac{1}{2}}{R+1}$, where $R$ is the
experimental Hong-Ou-Mandel enhancement. The mean value of the
experimental cloning fidelity for all our tests reported in Table I,
given by $\bar{F}_{exp}=(0.803\pm 0.008)$, is indeed
consistent with the predicted value ${F_{th}}=(0.805\pm 0.007)$.

The experimental coincidence rate $C_{tot}$ can be compared
with the predicted one, as determined from $C_{source}$ after taking
into account three main loss factors: state preparation probability
$p_{prep}$, successful cloning probability $p_{clon}$, and detection
probability $p_{det}$. $p_{prep}$ depends on the conversion
efficiency of the q-plate and on the probabilistic
efficiency of the quantum transferrer $\pi\rightarrow o_{2}$ (0.5),
thus leading to $p_{prep}=0.40\pm0.03$. For ideal input photon
states, the experimental success probability of the cloning process
on a single BS output mode is expected to be essentially equal to
the theoretical one $p_{clon}=3/8$. The probability $p_{det}$
depends on the q-plate and transferrer efficiencies
($0.8\times0.5$) plus the fiber coupling efficiency ($0.15-0.25$). Hence we have $p_{det} = 0.06-0.10$.
Therefore, the expected event rate is $C_{source} \times p_{prep}^2
\times p_{clon} \times p_{det}^2 \times \frac{1}{2} = 0.5-1.5$ Hz,
where the final factor $1/2$ takes into account the probability that
the two photons are split into different output modes of the
analysis FBS. Typically we had around 400 counts in 600s, consistent
with the expectations.

In the present experiment, for practical reasons, we adopted a post-selection technique
to identify when two photons emerge from the same output mode. In principle, post-selection could be replaced by
quantum non-demolition measurements.\\

\textbf{Generalization of the cloning process to dimension $d$.}
Let us assume that a photon in the unknown input d-dimensional state
$\ket{\varphi}$ to be cloned is injected in one arm of a balanced
beam splitter (BS), while the ``ancilla'' photon in the other arm is
taken to be in a fully mixed state
$\rho=\frac{I_d}{d}=\frac{1}{d}\sum_n{\ket{n}\bra{n}}$, where
$\ket{n}$ with $n=1,...,d$ is a orthonormal basis. Without loss of
generality, we may choose a basis for which
$\ket{1}\equiv\ket{\varphi}$. Depending on the state of the ancilla,
we must distinguish two cases in the input: (i) the two-photon state
is $\ket{1}\ket{1}$, with probability $1/d$, or (ii) it is
$\ket{1}\ket{k}$ and $k\neq 1$, with probability $(d-1)/d$. After
the interaction in the BS, we consider only the case of two photon
emerging in the same BS output mode (case of successful cloning).
Then, quantum interference leads to a doubled probability for case
(i) than for case (ii), so that the output probabilities are
respectively rescaled to $2/(d+1)$ for case (i) and $(d-1)/(d+1)$
for case (ii). The cloning fidelity is $1$ for case (i) and $1/2$
for case (ii), so that an overall fidelity of
$F=\frac{2}{d+1}\times1+\frac{(d-1)}{d+1}\times\frac{1}{2}=\frac{1}{2}+\frac{1}{d+1}$
is obtained, corresponding to the optimal one, as shown in Ref.\
\onlinecite{Nave03}. The success rate of the cloning is
$p=\frac{(d+1)}{2d}$.

\end{footnotesize}

\section*{Author Contribution}
E.N., L.S., F.S., F.D.M., L.M., E.S.: Conceived and designed the experiments; E.N., L.S., F.S.: Performed the experiments; E.N., L.S., F.S.: Analyzed the data; L.M., B.P., E.K., E.S.: Contributed materials; E.N., L.S., F.S., F.D.M., L.M., B.P., E.K., E.S.: Paper writing.

\end{document}